# Integrated Strategy for Urban Traffic Optimization: Prediction, Adaptive Signal Control, and Distributed Communication via Messaging

**ISMAIL ZRIGUI[1], SAMIRA KHOULJI[1], MOHAMED LARBI KERKEB[2]**

[1] Innovate Systems Engineering Lab(ISI), National School of Applied Sciences, Abdelmalek Essaadi University, Tetouan, Morocco

[2] Ibn Tofail University, Kenitra, Morocco

E-mail: [1]izrigui@uae.ac.ma

**ABSTRACT**

This work introduces an integrated approach to optimizing urban traffic by combining predictive modeling of vehicle flow, adaptive traffic signal control, and a modular integration architecture through distributed messaging. Using real-time data from various sensors, the system anticipates traffic fluctuations and dynamically adjusts signal phase durations to minimize delays and improve traffic flow. This proactive adjustment, supported by algorithms inspired by simulated annealing and reinforcement learning, also enhances energy efficiency, reduces pollutant emissions, and responds effectively to unexpected events (adverse weather, accidents, or temporary gatherings).

Preliminary simulations conducted in a realistic urban environment demonstrate a significant reduction in average waiting times. Future developments include incorporating data from connected vehicles, integrating new modes of transport, and continuously refining predictive models to address the growing challenges of urban mobility.

## 1. Introduction and Study Objectives:

Modern urban areas face an ongoing increase in traffic volume due to sustained population growth, accelerated urbanization, and a growing reliance on personal vehicles. This results in chronic congestion, longer travel times, and deteriorating air quality and living standards for residents [1]. Conventional traffic signal control, often based on fixed or semi-adaptive plans, struggles to cope with rapid and unpredictable traffic fluctuations. Without dynamic adjustments to real-time demand, the efficiency of the transport network is compromised, exacerbating congestion and long queues [2].

Emerging approaches aim to address these challenges by combining predictive modeling, algorithmic optimization, and real-time communication. Intelligent Transportation Systems (ITS) leverage diverse sensor data (inductive loops, cameras, environmental sensors) and predictive model capabilities to continuously adjust control parameters, such as signal phase durations [3]. These efforts aim to improve traffic flow, reduce emissions and energy consumption, and enhance resilience to unforeseen events (e.g., extreme weather or accidents).

The development of distributed and modular architectures, facilitated by message-oriented middleware (MOM), plays a crucial role in this evolution. These infrastructures enable the flexible and scalable integration of heterogeneous components: sensors, prediction modules, optimization systems, local signal controllers, and third-party services (e.g., fleet management or autonomous vehicles) [4]. Using an asynchronous messaging bus allows smooth information flow without major bottlenecks and supports adding or updating components without disrupting the entire system.

This study proposes an integrated approach combining near-real-time vehicle flow prediction, adaptive optimization of signal phase durations, and systemic orchestration through distributed messaging. Leveraging algorithms inspired by simulated annealing and reinforcement learning, the solution seeks to minimize intersection waiting times, improve overall traffic flow, and offer high evolutionary flexibility. Preliminary simulations conducted in a realistic urban environment reveal promising opportunities for large-scale implementation in contexts where intelligent traffic management is essential for smart city development.

## 2. State of the Art:

Traditional urban traffic management strategies primarily rely on fixed or semi-adaptive signal plans, characterized by predefined cycles for typical scenarios, such as rush hours. While these configurations have been the norm for many years, they struggle to adapt to unexpected situations, such as accidents or adverse weather conditions. Research [1] highlights the lack of responsiveness of these traditional strategies, fostering the emergence of more dynamic methods. Heuristic algorithms, such as genetic algorithms, have demonstrated their ability to reduce intersection delays under high congestion levels [3]. However, implementing these solutions requires significant computational resources, limiting their scalability.

**Advanced Techniques:**

a. **Reinforcement Learning**: Reinforcement learning offers adaptive traffic signal management, where the system continuously adjusts signals based on traffic conditions. This approach significantly improves urban mobility and reduces average stop durations, as shown in applied studies [4].
b. **Multi-objective Optimization**: Beyond reducing waiting times, environmental objectives like minimizing energy consumption and pollutant emissions have emerged as complementary goals [5]. Vehicle-to-infrastructure (V2I) communications enable the simultaneous consideration of multiple indicators, enhancing both operational performance and sustainability.
c. **V2I Technologies and Simultaneous Coordination**: Real-time information exchange between vehicles and infrastructure allows fine synchronization of signal phases and vehicle trajectories, optimizing traffic flow and reducing waiting times and energy consumption.
d. **Predictive Models**: Predictive approaches anticipate short-term conditions by integrating historical data, contextual information (weather, events), and real-time flows, enabling proactive adjustments to signal phase distributions [6].

Despite advancements, challenges remain:

a. **Computational Complexity**: Recent approaches often demand high computational power, challenging their application in dense, dynamic environments [7].

b. **Data Dependency**: Prediction accuracy and adjustments heavily depend on data quality. Noisy or incomplete data can hinder model performance, requiring enhanced robustness against uncertainties [8].
c. **Limited Interoperability**: Integrating distributed, heterogeneous systems remains complex, constraining large-scale deployment and underutilizing resources.

Recent innovations, such as reinforcement learning, multi-objective optimization, and V2I communications, provide unprecedented opportunities for modernizing urban traffic management. Addressing constraints related to computational complexity, uncertain data, and component interoperability is essential for widespread adoption. Further research is crucial to designing systems that harmoniously integrate intelligence, sustainability, and resilience in urban environments.

## 3. System Architecture :

The designed architecture for urban traffic optimization relies on a modular and distributed organization structured around four main categories of components: sensors (S), controllers (C), the prediction server (P), and the messaging infrastructure (M). This approach provides a solid foundation for flow forecasting, dynamic adjustment of signal durations, and the flexible integration of third-party services.

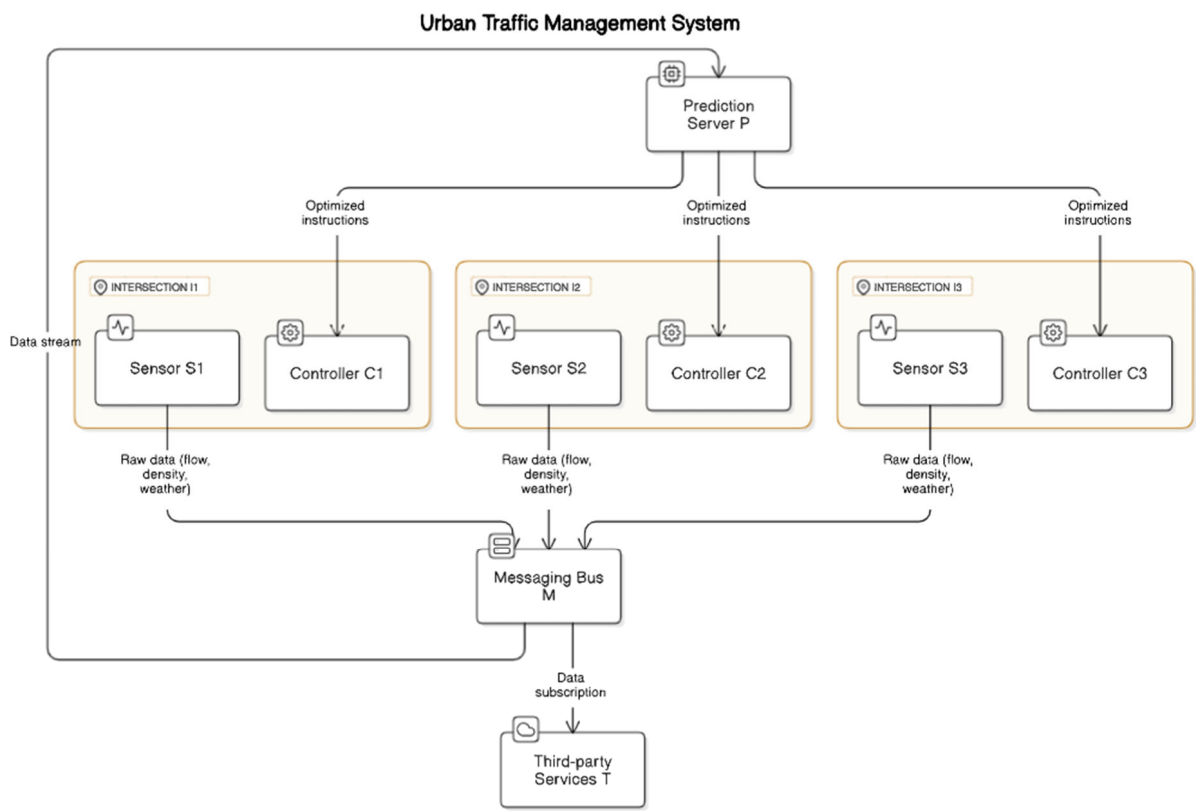

*Figure 1 : Conceptual Architecture Overview:*

This representation highlights the components and their interactions. Multiple urban intersections (I1, I2, I3) are equipped with sensors (S) and a local controller (C). Raw data (vehicle flow, densities, weather conditions) is transmitted via the messaging infrastructure (M) to the prediction server (P). The server returns optimized instructions to be applied at traffic lights, while third-party services (T) – such as fleet management systems – can subscribe to real-time information streams.

**Sensors (S):**

Sensors provide real-time data essential for assessing network conditions and anticipating traffic variations. These include:

- **Cameras:** Identify vehicle flow and detect unusual events.
- **Inductive Loops:** Measure vehicle counts integrated into the road surface.
- **Radars:** Record speeds to estimate traffic fluidity.
- **Weather Stations:** Provide visibility, precipitation, and other environmental data to improve forecasting accuracy.

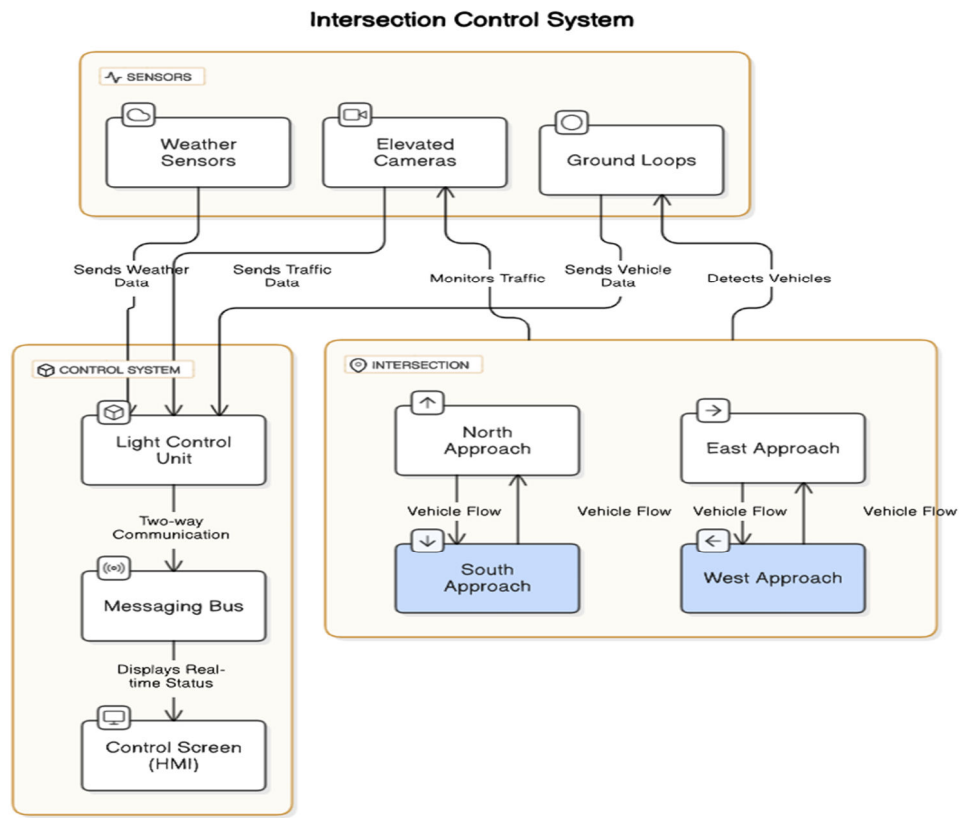

*Figure 2: Detailed View of a Connected Intersection*

This close-up of a single intersection showcases the variety of sensors (cameras, ground loops, weather sensors), the local controller responsible for managing the traffic lights, and a human-machine interface (HMI) that allows operators to monitor the situation in real time. The messaging bus (M) ensures bidirectional information transmission, guaranteeing seamless communication between the different components.

**Controllers (C):**

Installed in traffic light cabinets, controllers implement decisions derived from predictions and optimizations:

- Receive instructions via messaging, specifying green, red, or intermediate phase durations.
- Execute directives immediately, serving as the operational link between theoretical models and practical applications, ensuring reactive adaptation to traffic conditions.

**Prediction Server (P):**

The prediction server is the analytical backbone of the system:

- Leverages mathematical models and statistical learning techniques to estimate short-term flows (5-10 minutes).
- Integrates contextual data such as weather and event information to deliver reliable forecasts and contribute to cycle optimization.

**Messaging Infrastructure (M):**

The asynchronous messaging bus (e.g., MQTT or Kafka) facilitates seamless information exchange without bottlenecks or congestion:

- Components publish or subscribe to relevant data streams (traffic data, optimized instructions, alerts, etc.).
- Adding or removing services, such as a dedicated bus fleet management module or an intelligent navigation application, can be done without disrupting the system.

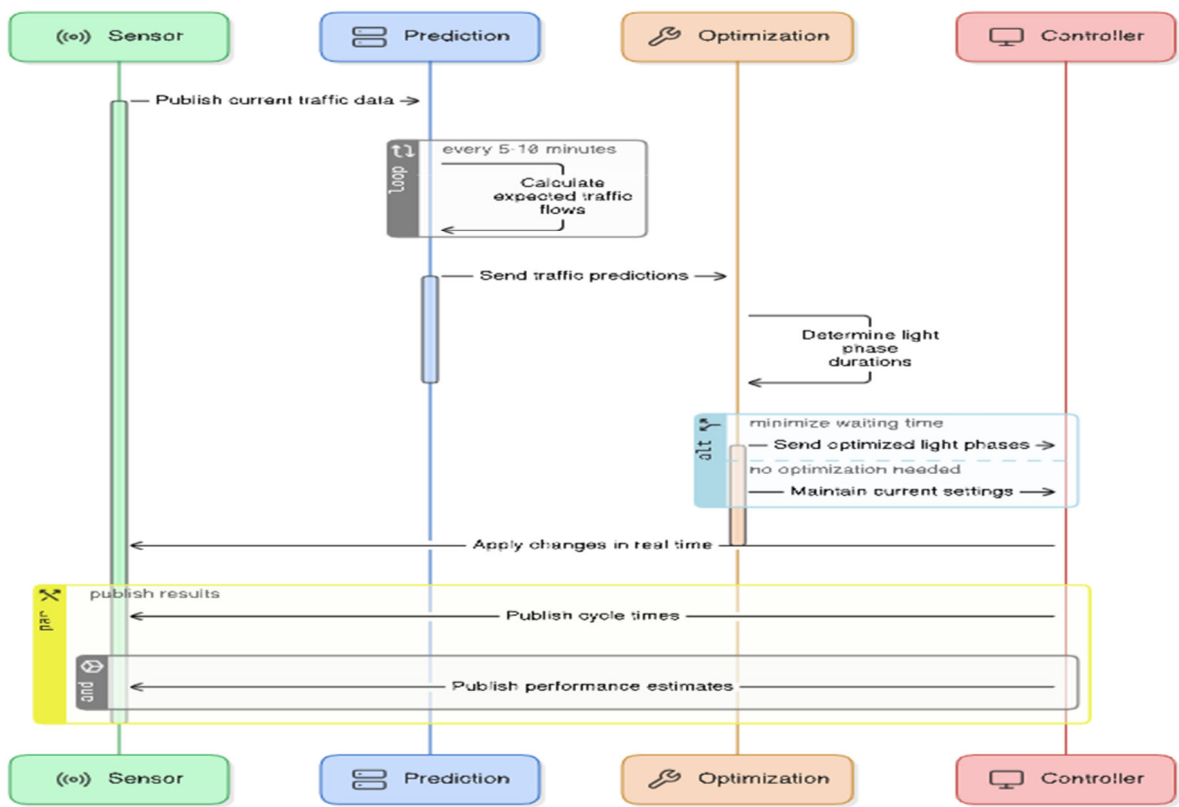

*Figure 3: Information Flow between Prediction, Optimization, and Control*

The sequence diagram clearly illustrates the process:

   a. Sensors publish their current traffic data.
   b. The prediction server calculates the expected traffic flows over a short time horizon.
   c. The optimization module—previously not represented but potentially integrated into the prediction server or functioning as an auxiliary component—determines the traffic light phase settings to minimize waiting times.
   d. The controller immediately implements these adjustments.
   e. Finally, the system publishes the results (cycle durations, performance indicators), making them available to third-party services and operators.

This integrated architecture combines prediction, local execution, and asynchronous communication. It adapts dynamically to urban conditions, enhancing traffic fluidity, reducing waiting times, and supporting the integration of new services. Its modularity and robustness make it a flexible and sustainable tool in building smarter and more sustainable mobility systems.

## 4. Modeling

This section outlines a mathematical framework designed to characterize the optimization problem for managing traffic signal timing at an urban intersection. The goal is to determine the optimal distribution of green phase durations, accounting for fluctuating traffic demand, operational constraints, and the integration of contextual data (e.g., weather, events, and network conditions). The overarching objective is to enhance traffic flow, reduce waiting times, and contribute to more sustainable mobility in complex urban environments [9,10].

**Conceptual Framework:**

Consider an intersection with N distinct approaches, denoted as i ∈ {1, 2, …, N}, each controlled by a traffic signal. The operation of these signals follows a complete cycle of duration C, divided into segments corresponding to green phases allocated to each approach. The decision variable $x_i(t)$ represents the green phase duration assigned to approach i during the cycle starting at time t. The control horizon T refers to the period over which the optimization of signal timing is applied, typically spanning a few minutes or several successive cycles.

**Objective Function:**

The primary evaluation criterion is the average waiting time of vehicles at the intersection. For each approach i, a function $w_i(t)$ estimates the average waiting time of users based on the flow $\lambda_i(t)$ and the green phase duration $x_i(t)$. The global optimization function aims to minimize the aggregated average waiting time across all approaches:

$$\min_{\{x_i(t)\}} W(t) = \frac{1}{N} \sum_{i=1}^{N} w_i(t)$$

where w_i(t) depends on the interplay between traffic flow and signal timing. This objective function is crucial for achieving efficient traffic management while balancing operational requirements and constraints.

The functions w_i(t) can be developed using theoretical queueing models or microsimulations calibrated with real-world data. Approaches inspired by traffic flow theory, such as the "Cell Transmission Model" [11], or statistical estimation methods based on empirical data, can be utilized to characterize the relationship between green time, flow, and waiting time.

**Operational Constraints:**

Three categories of constraints are essential to ensure the consistency and reliability of signal timing plans:

1. **Minimum and Maximum Green Phase Durations**: To ensure safety and signal clarity for drivers, lower and upper bounds are imposed on the duration of each phase:

   $$x_i^{\min} \leq x_i(t) \leq x_i^{\max}$$

   Such limits are commonly recommended in the literature to prevent undesirable situations, such as excessively short green phases that are difficult to perceive or overly long ones that disadvantage other approaches [12].

2. **Cycle Length Conservation**: The sum of all green phase durations, including possible interphase durations, must equal the cycle length:

   $$\sum_{i=1}^{N} x_i(t) = C$$

   This constraint ensures fundamental temporal consistency, allowing integration into a global control strategy, potentially coordinated with other intersections [13].

3. **Traffic Flow Prediction**: The incoming flow on each approach, λ_i(t), is not constant. It varies with time, weather, local events (e.g., gatherings, games, or construction), and the overall network state. λ_i(t) can be modeled as a relationship incorporating p temporal lags (past flow values) and q external variables (Z_j(t)):

   $$\lambda_i(t) = \alpha_i + \sum_{k=1}^{p} \beta_{i,k} \lambda_i(t-k) + \sum_{j=1}^{q} \gamma_{i,j} Z_j(t)$$

   The coefficients, and are typically determined through regression methods or machine learning. Accurate flow prediction is a major advantage, allowing proactive adjustment of green phase durations based on upcoming conditions rather than reacting late to congestion [14].

**Key Parameters:**

- $\lambda_i(t)$ : Estimated vehicle flow (vehicles per time interval) on approach i.
- $x_i(t)$: Green phase duration on approach i for the cycle starting at time t (seconds).
- $w_i(t)$: Modeled average waiting time on approach i, as a function of flow and green time.
- $Z_j(t)$: External variables (e.g., weather conditions, events, or information from connected vehicles).
- p,q: Number of temporal lags and additional variables integrated into the model.
- C: Total cycle length (seconds).
- $x_i^{\min}, x_i^{\max}$ : Imposed bounds on green phase durations.

This modeling framework serves as the foundation for applying advanced optimization methods, including stochastic approaches (e.g., simulated annealing or evolutionary algorithms) or reinforcement learning techniques. These allow continuous adjustments of phase distributions, integrating predictive models and constraints into a global optimization strategy. Such an approach enables adaptive management, capable of rapidly responding to demand variations and significantly improving network performance [15,16].

## 5. Optimization and Adaptive Control

Once the problem is modeled, the critical step involves determining the optimal green phase durations for each traffic cycle. The challenge lies in dynamically adapting to fluctuating and uncertain traffic, considering flow predictions, operational constraints, and external conditions. This problem falls under multivariate optimization, potentially non-convex, and therefore requires robust and flexible algorithms capable of delivering satisfactory solutions within limited computation time [17,18].

**General Optimization Framework:**

The goal is to identify the configuration that minimizes the objective function, typically average waiting time, while satisfying the previously defined constraints (bounds on durations, cycle conservation, etc.). At each control cycle , the system uses predicted flows and contextual parameters to adjust signal durations accordingly. Iterative methods are commonly employed to progressively improve an initial solution.

**Optimization Algorithm:**

1. **Data Input:** Predicted flows , constraints on , cycle length , and model parameters are provided as input.
2. **Initialization:** An initial solution is chosen, such as a uniform green time distribution across all approaches, or a pre-existing signal plan.
3. **Iteration Loop:**
   - **Performance Evaluation:** At each iteration, a cost function is calculated, representing the total waiting time or a multi-objective combination of waiting time, pollutant emissions, etc.
   - **Solution Improvement:** The values are adjusted using an appropriate optimization method. Possible techniques include:
     - Gradient-based approaches or numerical methods derived from nonlinear programming [19].

- Heuristics and metaheuristics, such as genetic algorithms [18], simulated annealing, or ant colony algorithms, particularly useful for complex combinatorial problems [20].
- Reinforcement learning approaches, where the agent "learns" from experience without an explicit model, gradually optimizing its decisions [21].
    - **Projection onto Feasible Space:** After generating a new candidate solution, constraints are verified. If necessary, projections or corrections are applied to ensure the sum of durations equals and each green phase duration respects the imposed bounds.
4. **Convergence Criterion:** The iterative process continues until a stopping criterion is met, such as a maximum number of iterations, negligible improvement in waiting time, or a computational time limit set for operational needs [17].
5. **Output:** The final solution is transmitted to the local controller, which immediately implements the adjusted signal durations for the next cycle. This procedure repeats at each reevaluation period, ensuring continuous adaptation to traffic conditions.

**Example: Simulated Annealing (Pseudocode):**

Simulated annealing is a stochastic optimization technique inspired by the metal cooling process. It is particularly suited to problems with multiple local minima [20]. The principle involves exploring the solution space while occasionally accepting less optimal solutions to avoid becoming trapped in a local minimum too early.

```
Input : Lambda, X_min, X_max, C, T_max, initial_solution
X = initial_solution
best_X = X
best_cost = Cost(X)

For T from T_{max} to a low temperature:
    X_new = Perturb(X)
    If FeasibilityCheck(X_new):
        Cost_new = Cost(X_new)
        Delta = Cost_new - Best_Cost
        If (Delta < 0) or (exp(-Delta / T) > Random(0,1)):
            X = X_new
            If Cost_new < Best_Cost:
                Best_X = X_new
                Best_Cost = Cost_new

Return Best_X
```

**Algorithm Parameters:**

- **Cost(X) :** Objective function, often the sum of average waiting times across all approaches.
- **Perturb(X) :** Operator applying a slight mutation to solution , such as modifying an axis' green time by a few seconds.
- **T_max :** Initial temperature controlling the probability of temporarily accepting less optimal solutions.

- **FeasibilityCheck(X_new) :** Ensures constraints are satisfied (e.g., sum equals , durations within bounds).

**Real-Time Adaptation and Control:**

One of the main advantages of these optimization approaches is their integration into real-time control systems, allowing frequent reevaluation of phase durations as traffic evolves. Recent studies demonstrate that adaptive strategies significantly reduce travel times and congestion compared to static approaches [17,21]. This adaptability is essential in constantly changing urban contexts, where traffic demand and external conditions (e.g., weather, events, incidents) vary rapidly.

## 6. Case Studies

To evaluate the relevance and effectiveness of the proposed modeling, prediction, and optimization approaches, it is essential to apply them to concrete scenarios. Case studies test the robustness of models and algorithms in realistic contexts, accounting for variations in traffic demand, weather conditions, and specific events that can drastically alter vehicle flows [22,23].

**Urban Intersection Scenario:**

Consider an intersection with four approaches integrated into a high-density urban network. This intersection is equipped with inductive loops and cameras providing real-time flow data. The predictive system, based on the previously described statistical models, generates traffic estimates for the next 5 to 10 minutes [24].

**Example of Realistic Conditions (Fictional):**

- **Observation Period:** Afternoon rush hour (5:00 PM–7:00 PM), when traffic volume peaks.
- **Weather Conditions:** Moderate rainfall (5 mm/h), reducing visibility and potentially increasing braking distances.
- **Local Event:** A football match nearby increases vehicle flow on one axis starting at 5:05 PM. Such events can significantly disrupt usual flow patterns and require a rapid response from the control system [25].

For illustration, suppose data collected at 5:00 PM indicates observed flows of approximately [300, 250, 180, 210] vehicles per 5 minutes on the four approaches (illustrative values). Based on this information, the predictive models incorporate weather, the sporting event, and past trends to estimate, at 5:05 PM, an increase in flow on the first axis (from 300 to 340 vehicles/5 min), while the other approaches experience more modest variations, such as [260, 200, 220] vehicles/5 min.

**Application of Optimization:**

The optimization algorithm, integrating a dynamic adjustment scheme for green phase durations, allocates a few extra seconds of "green" to the most loaded axis to reduce overall waiting time. In this case, it may involve adding 5 additional seconds to the green phase of the most congested approach, while redistributing durations across less critical axes to avoid unduly lengthening the total cycle. Preliminary results suggest a reduction in average waiting time and improved flow despite degraded conditions [26].

**Evaluation via Microsimulation:**

To validate these adjustments, traffic microsimulation tools such as VISSIM or Aimsun are used. These software solutions model vehicle behavior in detail and measure the impact of adaptive control strategies on various performance indicators (waiting time, travel time, emissions, fuel consumption). Comparisons between a fixed strategy (no adaptation), a semi-adaptive strategy, and the proposed method generally show significant improvements in congestion reduction and resilience to traffic disruptions [23,27].

Case studies thus provide concrete insights into how the proposed approaches integrate into real-world contexts. They help identify potential practical issues, such as sensitivity to data quality, the need for rapid updates, or the impact of green phase adjustments on soft modes of transport (pedestrians, cyclists). Lessons learned from these studies feed into future improvements at both the predictive model and optimization algorithm levels, paving the way for large-scale implementation in complex urban environments.

## 7. Integration via Messaging

Integrating the various components of the system (sensors, prediction modules, optimization algorithms, traffic light controllers, and third-party services) requires a communication infrastructure that is flexible, scalable, and resilient. In this context, adopting a "Message-Oriented Middleware" (MOM) solution is a relevant choice. MOM ensures asynchronous data exchange between the system's distributed entities without creating strong dependencies or bottlenecks. This approach simplifies adding or updating new modules, thereby facilitating the system's evolution and maintenance in a constantly changing urban environment [28,29].

**Principles of Messaging Communication:**

Instead of establishing direct connections between each component, the MOM infrastructure offers a centralized communication channel in the form of topics. System entities choose to publish data on certain topics or subscribe to receive it. This logic allows information producers (e.g., sensors) and consumers (e.g., the prediction server, optimization modules, or third-party services) to operate independently. Changes made to a component (e.g., changing data providers or adding a fault detection module) generally do not require major reconfiguration of the whole system as long as the messaging topics remain consistent.

**Examples of Information Flows:**

- **Sensors → traffic_flows/axisX:** Raw data on traffic intensity (vehicle counts, average speeds).
- **Prediction → predicted_flows/axisX:** Short-term flow estimates for axis X, based on historical, contextual, and real-time data.
- **Optimization → traffic_signal_decisions/intersectionY:** Instructions for adjusting green phase durations at intersection Y, determined by the optimization algorithm.
- **Controller → traffic_signal_status/intersectionY:** Confirmation that the local controller at the intersection has applied the transmitted settings, accompanied by a status of the current cycle's execution.

- **Third-Party Services → display, alerts:** Information dissemination to end-users (drivers, operators, fleet managers), for instance, via mobile applications or dynamic display panels. These services can also include external data providers, such as weather systems or event managers.

This modular architecture makes it easy to integrate new components. For example, a service dedicated to autonomous vehicles could subscribe to relevant information (predicted_flows, traffic_signal_status) to automatically adjust routes, reduce travel time, and enhance safety [30].

**Conceptual Diagram:**

A conceptual diagram (Figure 4) illustrates the architecture via a messaging bus. The various blocks (sensors, prediction, optimization, controllers, third-party services) communicate through a central messaging bus, represented by the gray cloud, where data flows are organized by topics.

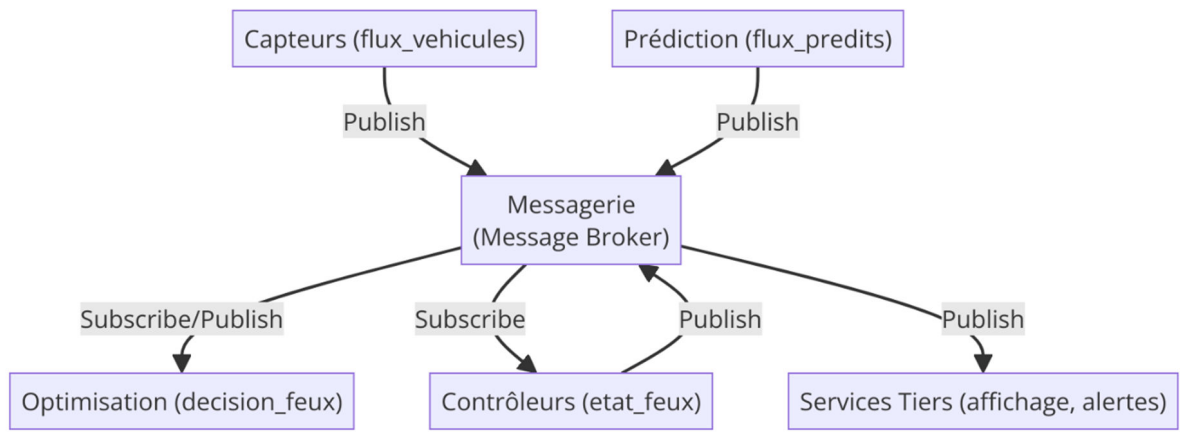

**Figure 4: Conceptual diagram of integration via a messaging bus.**

The messaging bus acts as an information hub between sensors, prediction modules, optimization algorithms, local traffic controllers, and third-party services. Each component can publish or subscribe to a set of defined topics (traffic_flows, predicted_flows, traffic_signal_decisions, traffic_signal_status, etc.), ensuring asynchronous and flexible communication.

With this architecture, the system can evolve without disrupting the entire network. Adding a new sensor, replacing an optimization module with a more efficient one, or integrating an external service becomes a straightforward operation limited to configuring topic subscriptions and publications.

## 8. Results and Discussion

Evaluations conducted on simulated scenarios using traffic microsimulation tools such as SUMO or VISSIM demonstrate the value of the presented approaches. Comparing the integrated adaptive strategies (including flow prediction, dynamic signal optimization, and messaging bus use) with static or semi-adaptive configurations reveals notable improvements in performance indicators.

Across a set of representative case studies for a dense urban network, simulations show an average reduction of approximately 12% in waiting times during peak hours. This reduction results in smoother traffic flow, increased user comfort, and better utilization of existing infrastructure. Additionally, as vehicles spend less time idling, pollutant emissions can drop by approximately 8% under the same conditions, contributing to sustainable mobility goals and a reduced environmental footprint [31,32].

One major advantage of using an asynchronous messaging architecture is the flexibility it offers to integrate new services without disrupting the entire system. For instance, introducing a weather forecasting module can further refine prediction quality and anticipate atmospheric conditions' impact on traffic. This update is achieved simply by adjusting subscriptions to relevant topics without deeply altering the software structure. Simulation results indicate that this integration enhances robustness against uncertainties, enabling optimization algorithms to maintain stable performance even during sudden changes, such as unexpected precipitation or reduced visibility [33].

Moreover, the modularity of the approach facilitates the future addition of other information sources, such as data from connected vehicles, signals from fleet management systems, or information related to soft mobility modes (pedestrians, cyclists). This extensibility prepares the ground for truly multimodal and resilient traffic management, capable of adapting to the increasing diversity of urban environments and mobility practices [34].

Overall, the obtained results underline the relevance of the methodological choices (prediction, adaptive optimization, modular messaging integration). They pave the way for broader operational deployments and complementary studies aimed at refining models, reducing computational costs, or integrating new objectives (e.g., improving safety, enhancing public transport efficiency, etc.).

## 9. Conclusion and Perspectives

This study has illustrated an integrated approach to optimizing urban traffic by leveraging the synergy between predictive modeling of flows, adaptive traffic signal control, and modular communication infrastructure based on messaging. Preliminary results, obtained from simulations grounded in realistic data and scenarios, are encouraging. They highlight significant reductions in waiting times, decreased pollutant emissions, and enhanced resilience to variable and unpredictable traffic conditions.

Beyond these initial successes, numerous opportunities lie ahead. First, integrating data from connected vehicles (V2X) could significantly refine predictions and enable more responsive optimization strategies. Information exchange between vehicles and infrastructure allows for better signal synchronization, finer trajectory management, and even proactive coordination to minimize stops and unnecessary acceleration phases [30].

Second, adopting collaborative strategies across multiple intersections promises to extend the impact of the proposed approaches. By coordinating decisions at the corridor or broader network level, it becomes possible to improve traffic flow not just locally but across entire neighborhoods or even cities. Transitioning to distributed, multi-level optimization requires advanced data orchestration and algorithmic techniques but could lead to substantial gains in overall performance [35].

Finally, addressing alternative transportation modes, such as bicycles, electric scooters, or autonomous shuttles, is imperative to meet the demands of multimodal mobility. Adapting optimization algorithms to the diversity of users, including pedestrians, presents a significant methodological and technological challenge. This involves developing more comprehensive models that simultaneously integrate safety, comfort, sustainability, and equity among different user types [36,37].

In summary, the presented work provides a solid foundation for the development of intelligent, adaptive, and sustainable urban traffic management systems. The opportunities unlocked by technological advancements, the rise of connected data, and innovations in optimization and communication pave the way for a future where urban traffic will flow more smoothly, safely, and in a more environmentally friendly manner.